\documentclass[11pt]{article}
\usepackage{epsfig}
\newcommand{\sss}{\vspace{.2in}}

\newcommand{\be}{\begin{equation}}
\newcommand{\ee}{\end{equation}}
\newcommand{\bea}{\begin{eqnarray}}
\newcommand{\eea}{\end{eqnarray}}
\newcommand{\sn}{{\rm sn}}
\newcommand{\cn}{{\rm cn}}
\newcommand{\dn}{{\rm dn}}
\newcommand{\sech}{{\rm sech}}

\begin{document}
\vspace{.2in}
~\hfill{\footnotesize UICHEP-TH/01-6,~~\today}
\vspace{.5in}
\begin{center}
{\LARGE {\bf Linear Superposition in Nonlinear Equations}}
\end{center}
\vspace{.5in}
\begin{center}
{\Large{\bf  \mbox{Avinash Khare}\footnote{Permanent address: Institute of Physics, Sachivalaya Marg, Bhubaneswar 751005, Orissa, India}  and
   \mbox{Uday Sukhatme} 
 }}
\end{center}
\vspace{.4in}
\noindent
{\large Department of Physics, University of Illinois at Chicago, Chicago, IL 60607-7059, U.S.A.} 
\\
 \\
\sss
\vspace{1.1in}
\begin{center}
{\Large {\bf Abstract}}
\end{center}
Even though the KdV and modified KdV equations are nonlinear, we show that suitable linear combinations of known periodic solutions involving Jacobi elliptic functions yield a large class of additional solutions. This
procedure works by virtue of some remarkable new identities satisfied by the elliptic functions.
\newpage

\sss

The Korteweg-de Vries (KdV) and modified Korteweg-de Vries (mKdV) equations are known to possess many remarkable properties. They are examples of completely integrable systems with soliton solutions which have diverse applications of physical interest. Soliton solutions of the KdV equation
\be\label{E1}
u_t-6uu_x+u_{xxx} = 0 ~,
\ee		
are discussed in many texts \cite{dra}. The simplest, periodic, cnoidal travelling wave 
solution is
\be\label{E2}
u_1(x,t) = -2 \alpha^2 \dn^2(\xi_1,m)+\beta \alpha^2~,~~\xi_1 \equiv \alpha(x-b_1\alpha^2 t)~,
\ee
where $\alpha, m$ and $\beta$ are constants, and the ``velocity" $b_1$ is given by
\be\label{E3}
b_1 = 8 - 4m -6\beta ~.
\ee			
Here, we use the standard notation $\dn \,(\xi,m),~\sn \,(\xi,m),~\cn \,(\xi,m)$ to denote Jacobi elliptic functions, where $m$ is the elliptic modulus parameter $(0 \le m \le 1)$. Note that the period of solution (2) is $2K(m)/\alpha$, where $K(m)$ is the complete elliptic integral of the first kind \cite{abr}. In the limiting case $m=1$, one recovers the familiar result $-2 \alpha^2 \sech^2(\alpha(x-b_1\alpha^2 t))$. The properties of periodic multisoliton solutions more general than eq. (2) have also been discussed \cite{nov}. 

In this paper, we make suitable linear combinations of solution (2) to obtain a large class of new periodic solutions of the nonlinear KdV equation. This procedure works by virtue of some remarkable identities satisfied by the Jacobi elliptic functions, which to the best of our knowledge,  do not seem to be previously known. Our solutions consist of adding terms of the kind given in (2) but centered at $p$ equally spaced points along the period $2K(m)/\alpha$, where $p$ is any integer. The $p$-point solution is
\be\label{E4}
u_p(x,t) = -2 \alpha^2 \sum_{i=1}^{p} d_i^2 + \beta \alpha^2~; ~ d_i \equiv \dn[\xi_p + \frac{2(i-1)K(m)}{p},m]~, ~\xi_p \equiv \alpha(x-b_p\alpha^2 t) ~.
\ee
Clearly, $p=1$ is the original solution, but for any other $p$, we have new solutions of period $2K(m)/p\alpha$. For convenience, we also define the quantities $s_i$ and $c_i$ in analogy to the quantity $d_i$ defined above:
\be\label{E5}
~ s_i \equiv \sn[\xi_p + \frac{2(i-1)K(m)}{p},m]~~,
~ c_i \equiv \cn[\xi_p + \frac{2(i-1)K(m)}{p},m]~~.
\ee 

To see why eq. (\ref{E4}) is a solution of the KdV equation (\ref{E1}), it is best to first consider several small values of $p$. 
For $p=2$, substitution of eq. (\ref{E4}) into the left hand side of the KdV equation (\ref{E1}) gives on simplification:
\be\label{E6}
4m\alpha^5\{(8 - 4m-b_2-6\beta)(s_1 c_1 d_1+ s_2 c_2 d_2)+12(d_1^2 s_2 c_2 d_2 + d_2^2 s_1 c_1 d_1)\}.
\ee
One gets a solution only if the above expression vanishes. For $p=2$, the quantities $d_1 = \dn(\xi_2,m)$ and $d_2 = \dn [\xi_2+K(m),m]$ are well-known \cite{abr} to satisfy  $d_1^2 d_2^2 = 1-m$, which on differentiation gives $d_1^2 s_2 c_2 d_2 + d_2^2 s_1 c_1 d_1=0.$ Therefore, $u_2(x,t)$ is a solution of the KdV equation with velocity $b_2= 8 - 4m-6\beta,$ the same velocity as $b_1$.

For $p=3$, substitution of eq. (\ref{E4}) into the left hand side of the KdV equation (\ref{E1}) leads to the expression:
\bea\label{E7}
&&4m\alpha^5\{(8 - 4m -b_3-6\beta)(s_1 c_1 d_1+ s_2 c_2 d_2 + s_3 c_3 d_3)\nonumber\\
&+&12[d_1^2 (s_2 c_2 d_2 + s_3 c_3 d_3)+ d_2^2 (s_3 c_3 d_3 + s_1 c_1 d_1)+d_3^2 (s_1 c_1 d_1 + s_2 c_2 d_2)] \}.
\eea
In this case, $d_1 = \dn(\xi_3,m), d_2 = \dn (\xi_3+2K(m)/3,m), d_3 = \dn (\xi_3+4K(m)/3,m).$ In contrast to the $p=2$ case, there are no well-known relations between Jacobi elliptic functions at the arguments chosen above. However, we have recently discovered 
numerous new identities \cite{kha} which reduce by 2 (or a larger even number) the degree of cyclic homogeneous polynomials in Jacobi elliptic functions. The relevant identity for the $p=3$ case is
\be\label{E8}
d_1^2 (s_2 c_2 d_2 + s_3 c_3 d_3)+ d_2^2 (s_3 c_3 d_3 + s_1 c_1 d_1)+d_3^2 (s_1 c_1 d_1 + s_2 c_2 d_2) = \left( 2-\frac{2}{Q} \right) (s_1 c_1 d_1+ s_2 c_2 d_2 + s_3 c_3 d_3)~,
\ee
where 
\be\label{E9}
Q \equiv \sn^2 [2K(m)/3,m]~.
\ee
Identity (\ref{E8}) can be established analytically using the addition theorems \cite{abr} for Jacobi elliptic functions. It is also easy to check numerically. Substituting identity (\ref{E8}) into expression (\ref{E7}), shows that $u_3(x,t)$ is a solution of the KdV equation with velocity
\be\label{E10}
b_3 = 8 - 4m-6\beta + 24\left( 1-\frac{1}{Q} \right) ~.
\ee
Similarly, the identity analogous to (\ref{E8}) for $p=4$ is
\be\label{E11}
\sum_{i=1}^{4} d_i^2 \sum_{j \ne i}^4 s_j c_j d_j = -2\sqrt{1-m} \sum_{i=1}^{4} s_i c_i d_i~,
\ee
and consequently, the expression for the velocity is
\be\label{E12}
b_4 = 8 - 4m-6\beta-24\sqrt{1-m} ~.
\ee
For an arbitrary integer $p$, we have found that the relevant identity is
\be\label{E13}
\sum_{i=1}^{p} d_i^2 \sum_{j \ne i}^p s_j c_j d_j = A(p,m) \sum_{i=1}^{p} s_i c_i d_i~,
\ee
where the constant $A(p,m)$ can be evaluated in general by choosing any specific convenient value of the argument $\xi$ of the Jacobi elliptic functions. The values at $m=0,1$ are particularly simple:
\be\label{E14}
A(p,m=0) = -\frac{1}{3}(p-1)(p-2)~;~	A(p,m=1) = 0~.
\ee
The general expression for the velocity of our $p$-point solution (\ref{E4}) is
\be\label{E15}
b_p = 8 - 4m-6\beta +12 A(p,m) ~.
\ee
In Figure 1, we plot the velocity $b_p$ as a function of the modulus parameter $m$ for various values of $p$. In making the plots, we have chosen $\beta = 0$. Note that for this choice of $\beta$, the velocity is always positive and corresponds to right moving waves for $p \le 3$, whereas the velocity changes sign for $p > 3$.                                                                     

Two types of mKdV equations are discussed in the literature \cite{dra} corresponding to the sign chosen for the nonlinear term. We will consider these two types of mKdV equations separately, in order to stay with real solutions. The first type of mKdV equation is
\be\label{E16}
v_t-6v^2v_x+v_{xxx} = 0 ~.
\ee
The best known periodic travelling wave solution is
\be\label{E17}
v_1(x,t) = \pm \sqrt{m} \alpha \, \sn(\eta_1,m)~,~~\eta_1 \equiv \alpha(x-q_1\alpha^2 t)~,
\ee
with ``velocity" $q_1 = -(1+m)$ and period $4K(m)/\alpha$. In the limiting case $m=1$, $v_1(x,t)$ reduces to the familiar form $\pm \alpha \tanh(\alpha(x+2\alpha^2 t))$. 

For any odd integer $p$, we find the following solutions of the mKdV equation (\ref{E16}) by a specific linear superposition of the basic solution (\ref{E17}):
\be\label{E18}
v_p(x,t) = \pm \sqrt{m} \alpha \sum_{i=1}^p {\tilde s}_i~,~~p={\rm odd}~,
\ee
where we define
\bea
&&{\tilde s}_i \equiv \sn[\eta_p+\frac{4(i-1)K(m)}{p},m]~;~{\tilde c}_i \equiv \cn[\eta_p+\frac{4(i-1)K(m)}{p},m]~;\nonumber \\
&&{\tilde d}_i \equiv \dn[\eta_p+\frac{4(i-1)K(m)}{p},m]~;~\eta_p \equiv \alpha(x-q_p\alpha^2 t)~.
\eea
In order to verify that $v_p(x,t)$ satisfies eq. (\ref{E16}), one needs the identity
\be\label{E19}
m\sum_{i=1}^{p} {\tilde s}_i^2 \sum_{j \ne i}^{p} {\tilde c}_j {\tilde d}_j + 2m \left[\sum_{i<j}^{p} {\tilde s}_i {\tilde s}_j \right]\left[\sum_{k=1}^{p} {\tilde c}_k {\tilde d}_k \right] = \{B(p,m)-C(p,m)\}\left[\sum_{k=1}^{p} {\tilde c}_k {\tilde d}_k \right]~,~p={\rm odd},
\ee
which can be established from the much simpler identities \cite{kha}
\be\label{E20}
m\sum_{i<j}^{p} {\tilde s}_i {\tilde s}_j = B(p,m)~;~ m\sum_{i<j<k}^{p} {\tilde s}_i {\tilde s}_j {\tilde s}_k = C(p,m)\sum_{i=1}^{p} {\tilde s}_i~~,~p={\rm odd},
\ee
by using the properties of Jacobi elliptic functions.
The general expression for the velocity is
\be\label{E21}
q_p = -(1+m)-6\{B(p,m)-C(p,m)\}~.
\ee
Some explicitly computed values of the constants $B(p,m)$ and $C(p,m)$ are:
$$B(1,m)=0~;~C(1,m)=0~;~B(3,m)= -mQ ~;~C(3,m)=-1/Q~.$$

{}For any even integer $p$, the linear superposition of elementary solutions does not work, since identity (\ref{E20}) is restricted to odd values of $p$. However, remarkably enough, for this case, we find that products of elementary solutions do give new solutions! For example, the solutions for $p=2$ and $p=4$ are:
\be\label{E22}
v_2(x,t)= \pm \alpha m s_1 s_2~;~v_4(x,t)= \pm \alpha m (1-\sqrt{1-m}) s_1 s_2 s_3 s_4~.
\ee
These solutions can be verified using the general identity \cite{kha}
\be\label{E23}
\sum_{i<j<k}^{p} \frac{c_i d_i c_j d_j c_k d_k}{s_i s_j s_k}  = D(p,m)\sum_{i=1}^{p} \frac{c_i d_i}{s_i} ~,~p={\rm even}~.
\ee
The expressions for the velocities are
\be\label{E24}
q_2=-2(2-m)~;~q_4= -2(2-m)-12 \sqrt{1-m}~.
\ee 

It is well-known \cite{dra} that if $v(x,t)$ is a solution of the mKdV equation (\ref{E16}), then the Miura transform $u(x,t) = v^2 \pm v_x$ is a solution of the KdV equation (\ref{E1}). Therefore, using solutions (\ref{E18}) and (\ref{E22}), we can obtain even more solutions of the KdV equation via the Miura transform.  The simplest solution $(p=1)$ coming from eq. (\ref{E18}) is 
\be\label{E25}
u(x,t)=\alpha^2[m\sn^2(\eta,m) \pm \sqrt{m}\cn(\eta,m)\dn(\eta,m)]~,
\ee
with velocity $-(1+m)$, corresponding to a left moving wave. In view of the identities (\ref{E20}), it is clear that for odd integer $p$, appropriate linear superposition of the simplest solution (\ref{E25}) also solves the KdV equation with the same velocity $q_p$ given in eq. (\ref{E21}). Similarly, the solution resulting from a Miura transform of $v_2(x,t)$ in eq. (\ref{E22}) is 
\be\label{E26}
u(x,t)=\alpha^2[m^2 s_1^2 s_2^2 \pm m( s_1 c_2 d_2 + s_2 c_1 d_1)]~.
\ee
However, this is not a new solution since it simplifies to give either $\alpha^2(2ms_1^2-m)$ or $\alpha^2(2ms_2^2-m)$, both of which are KdV solutions (\ref{E2}) with $\beta = 2-m$.

Finally, we consider the second type of mKdV equation
\be\label{E27}
v_t+6v^2v_x+v_{xxx} = 0 ~.
\ee
Here again, we have found several new solutions via linear superposition: 
\be
v_p(x,t) = \pm  \alpha \sum_{i=1}^p d_i~,~q_p = 2-m+6(E(p,m)-F(p,m))~,~ (p=1,2,3,...),
\ee 
\be
v_p(x,t) = \pm \sqrt{m}  \alpha \sum_{i=1}^p {\tilde c}_i~,~q_p = 2m-1+6(G(p,m)-H(p,m))~,~ p={\rm odd},
\ee 
\be
v_p(x,t) = \pm  \alpha \sum_{i~{\rm odd}}^{p-1} (d_i-d_{i+1})~,q_p = 2-m-6(I(p,m)-J(p,m)+L(p,m))~,~ p={\rm even},
\ee 
where the constants $E(p,m), F(p,m),$ etc. are defined by the identities \cite{kha}
\bea
&&\sum_{i < j}^p d_i d_j = E(p,m)~;~ \sum_{i<j<k}^p d_i d_j d_k= F(p,m) \sum_{i}^p d_i~;\nonumber \\
&&m\sum_{i < j}^p {\tilde c}_i {\tilde c}_j = G(p,m)~;~ m\sum_{i<j<k}^p {\tilde c}_i {\tilde c}_j {\tilde c}_k= H(p,m) \sum_{i}^p {\tilde c}_i~;\nonumber \\
&&\sum_{i < j}^p d_i d_j = I(p,m)~,~(i+j)~ {\rm odd}~~~;~~~\sum_{i < j}^p d_i d_j = J(p,m)~,~(i+j)~ {\rm even}~;\nonumber \\
&&\sum_{\stackrel{i<j<k}{i+j+k={\rm odd}}}^p d_i d_j d_k -\sum_{ \stackrel{i<j<k}{i+j+k={\rm even}}}^p d_i d_j d_k= L(p,m)\sum_{i~{\rm odd}}^{p-1} (d_i-d_{i+1})~.
\eea
Some explicit simple values at $m=0$ are:
\bea
&&E(p,0)= p(p-1)/2~,~F(p,0)=(p-1)(p-2)/6~,~G(p,0)=0~,~H(p,0)=(p^2-1)/6~,\nonumber \\
&&I(p,0)=p^2/4~,~J(p,0)=p(p-2)/4~,~L(p,0)=(p-1)(p-2)/6~.
\eea
At $m=1$ all constants $E,F,G,H,I,J,L$ vanish for any $p$. Also, the constants for arbitrary $m$ but specified small values of $p$ are:
\bea
&&E(2,m)=I(2,m)=\sqrt{1-m}~,~F(2,m)=J(2,m)=L(2,m)=0~,\nonumber \\
&&E(3,m)=1-mQ+2\sqrt{1-mQ}~,~F(3,m)=(1-Q)/Q~,\nonumber \\
&&G(3,m)=-m(1-m)Q/(1-mQ)~,~H(3,m)=(1-mQ)/Q~,\nonumber \\
&&E(4,m)=2r(1+r+r^2)~,~F(4,m)= L(4,m)=r^2~,\nonumber \\
&&I(4,m)=2r(1+r^2)~,~J(4,m)=2r^2~,~r \equiv (1-m)^{1/4}~.
\eea

Our method of judicious linear superposition for finding new solutions also works for many other nonlinear equations. Indeed, we have verified that it works for the nonlinear Schr{\" o}dinger and KP equations as well as the $\lambda \phi^4$ model. These results will be reported in a forthcoming publication \cite{coo}.

The solutions we have obtained for the KdV equation all correspond to one gap periodic potentials. This process can be generalized to obtain  periodic potentials with a finite number of band gaps \cite{ks,kha1}.

One of us (A.K) thanks the Department of Physics at the University of Illinois at Chicago for hospitality. We acknowledge grant support from the U.S. Department of Energy.

\noindent {\Large \bf Figure Caption}

\noindent {Figure 1: Velocity $b_p$ of the periodic KdV travelling wave solution $u_p(x,t)$ [eq. (\ref{E4})] as a function of the modulus parameter $m$ for $p=1,2,3,4.$}

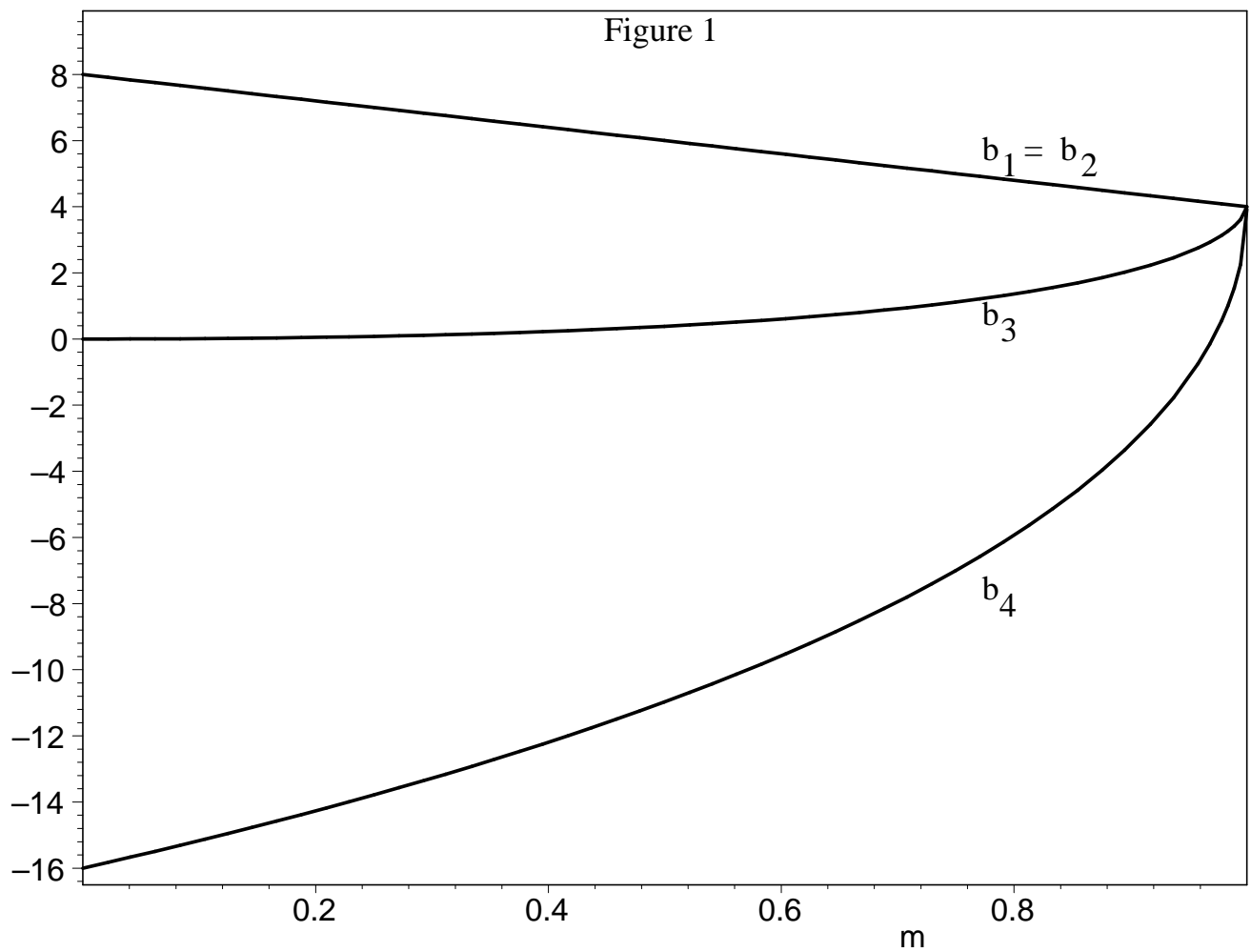
\end{document}